\documentclass[prd,twocolumn,showpacs,superscriptaddress]{revtex4}
\usepackage{amsfonts}
\usepackage{amsmath}
\usepackage{amssymb}
\usepackage{amsthm}
\usepackage{bm}
\usepackage{dcolumn}
\usepackage{epsfig}
\usepackage{graphicx}
\usepackage{graphics}
\usepackage[latin1]{inputenc}
\usepackage{latexsym}
\usepackage{rotating}
\usepackage{hyperref}
\usepackage{amsthm}
\usepackage{xcolor}
\newcommand{\beqs}{\begin{equation*}}
\newcommand{\beq}{\begin{equation}}

\newcommand{\eeqs}{\end{equation*}}
\newcommand{\eeq}{\end{equation}}

\newcommand{\beqas}{\begin{eqnarray*}}
\newcommand{\beqa}{\begin{eqnarray}}

\newcommand{\eeqas}{\end{eqnarray*}}
\newcommand{\eeqa}{\end{eqnarray}}

\newcommand{\eq}[2]{\begin{equation} #1 \label{#2} \end{equation}}

\newcommand{\al}{\alpha}

\newcommand{\ga}{\gamma}
\newcommand{\de}{\delta}

\newcommand{\La}{\Lambda}

\newcommand{\blist}{\begin{itemize}}

\newcommand{\elist}{\end{itemize}}

\providecommand{\href}[2]{#2}

\DeclareFontFamily{OT1}{rsfs}{}
\DeclareFontShape{OT1}{rsfs}{m}{n}{ <-7> rsfs5 <7-10> rsfs7 <10->rsfs10}{} 
\DeclareMathAlphabet{\mycal}{OT1}{rsfs}{m}{n}

\DeclareMathOperator{\extdm}{d}
\newcommand{\extd}{\extdm \!}

\begin{document}
\title{Cosmological constant as confining U(1) charge in two-dimensional dilaton gravity}

\author{Daniel Grumiller}
\email{grumil@hep.itp.tuwien.ac.at}
\affiliation{Institute for Theoretical Physics, Vienna University of Technology, Wiedner Hauptstrasse 8--10/136, A-1040 Vienna, Austria}

\author{Robert McNees}
\email{rmcnees@luc.edu}
\affiliation{Loyola University Chicago, Department of Physics, Chicago, IL 60660, USA}

\author{Jakob Salzer}
\email{salzer@hep.itp.tuwien.ac.at}
\affiliation{Institute for Theoretical Physics, Vienna University of Technology, Wiedner Hauptstrasse 8--10/136, A-1040 Vienna, Austria}

\date{\today}
\begin{abstract}
The cosmological constant is treated as a thermodynamical parameter in the framework of two-dimensional dilaton gravity. We find that the cosmological constant behaves as a $U(1)$ charge with a confining potential, and that such potentials require a novel Born--Infeld boundary term in the action. The free energy and other thermodynamical quantities of interest are derived, from first principles, in a way that is essentially model-independent. We discover that there is always a Schottky anomaly in the specific heat and explain its physical origin. Finally, we apply these results to specific examples, like Anti-de~Sitter--Schwarzschild--Tangherlini black holes, Ba\~nados--Teitelboim--Zanelli black holes and the Jackiw--Teitelboim model.
\end{abstract}
\maketitle

\section{Introduction}

About 30 years ago, Teitelboim and Henneaux proposed a mechanism to account for the small value of the cosmological constant \cite{Teitelboim:1985dp}, \cite{Henneaux:1985tv}: If four-dimensional gravity without cosmological constant is coupled to an antisymmetric gauge field $A_{\mu \nu \rho}$ the cosmological constant $\Lambda$ reappears as a constant of motion. Therefore, the cosmological constant may be considered to be dynamical, since different values of the constant of motion correspond to different values of $\Lambda$. One might conceivably find a mechanism that drives this dynamical cosmological constant to small values \cite{Brown:1988kg}.

Although it is not clear whether such a mechanism actually works in our Universe \cite{Bousso:2000xa}, it is interesting in its own right to consider the cosmological constant not as an input parameter of the action but instead as a state-dependent constant that may be changed \mbox{(thermo-)}dynamically. 
If one allows for negative values of the cosmological constant then novel aspects of the Anti-de~Sitter/conformal field theory (AdS/CFT) correspondence can be addressed where the length scale set by the AdS radius is no longer a fixed input parameter but becomes state-dependent.

Following this reasoning, many authors regarded the cosmological constant (mostly with negative sign) as a thermodynamic variable in black hole thermodynamics \cite{Sekiwa:2006qj,Dolan:2010ha,Dolan:2013ft,Dolan:2012jh,Dolan:2011xt,Cvetic:2010jb,Kastor:2009wy}. 
One of the most interesting results of these works was the insight that the thermodynamic variable conjugate to the cosmological constant is proportional to a negative volume. Since the cosmological constant may be regarded as a negative pressure the resulting first law contains a $V\extd p$ term. This implies that the mass of a black hole should correspond to the enthalpy of the system rather than the internal energy  \cite{Cvetic:2010jb,Kastor:2009wy,Larranaga:2011wd,Larranaga:2012pq,Altamirano:2014tva}. See also \cite{Kubiznak:2012wp,Gunasekaran:2012dq,Zhao:2013oza,Lu:2013hx,Johnson:2014yja,Johnson:2014xza} for more recent developments.

In this work we investigate the cosmological constant as a state-dependent parameter in two-dimensional (2d) dilaton gravity. Two-dimensional dilaton gravity is a rewarding model to study since it provides useful insights while not being as technically involved as higher dimensional theories. Some specific classes of solutions of higher dimensional theories can be studied in 2d dilaton gravity, including AdS-Schwarzschild--Tangherlini black holes and Ba\~nados--Teitelboim--Zanelli (BTZ) black holes. Some intrinsically 2d models are also interesting in their own right, like the Jackiw--Teitelboim model. For a general review of dilaton gravity see \cite{Grumiller:2002nm}. 

As will be shown, the cosmological constant may be treated in 2d dilaton gravity as a $U(1)$ charge with non-minimal coupling. This effectively leads to a confining electrostatic potential. The variation of the cosmological constant appears in the first law of black hole thermodynamics for the same reason the variation of charge would appear in the first law for a charged (e.g.~Reissner--Nordstr\"om) black hole. Thus, the cosmological constant emerges as a thermodynamic variable naturally in this treatment.

One of our main results is a novel Born--Infeld type of boundary term $I_{\textrm{\tiny BI}} $ that has to be added to the action, 
\eq{
\Gamma[g_{\mu\nu},\, A_\mu,\, X] = I_{\textrm{\tiny bulk}} + I_{\textrm{\tiny GHY}} + I_{\textrm{\tiny BI}} 
}{eq:intro1}
with
\begin{align}
 I_{\textrm{\tiny bulk}} &= -\tfrac{1}{2}\int_{\mathcal{M}}\!\!\!\extd^2x \sqrt{g} \,\left(XR-U(\nabla X)^2-2V\right) \nonumber\\ &\!\!\!\!\!\!\!\!\!\!\!\! +\int_{\mathcal{M}}\!\!\!\extd^2x\sqrt{g}\,fF^{\mu \nu}F_{\mu\nu} -4\int_{\mathcal{M}}\!\!\!\extd^2x \sqrt{g}\,\nabla_{\mu}(fF^{\mu \nu}A_{\nu}) \label{eq:intro1.5} \\ 
 I_{\textrm{\tiny GHY}} &= -\int_{\partial\mathcal{M}}\!\!\!\!\!\extd x\sqrt{\gamma}\,XK \\
 I_{\textrm{\tiny BI}} &= \int_{\partial \mathcal{M}}\!\!\!\!\!\extd x \sqrt{\gamma} \,\sqrt{e^{-Q}\left(w+2 F_{\mu \nu}F^{\mu \nu}f^2h\right)}\,.
\label{eq:intro2}
\end{align}
The bulk action $I_{\textrm{\tiny bulk}}$ depends on three arbitrary functions of the dilaton field $X$, namely on the kinetic potential $U$, the dilaton self-interaction potential $V$, and the gauge coupling function $f$. The boundary term $I_{\textrm{\tiny GHY}}$ is the dilaton analogue of the Gibbons--Hawking--York boundary term, which is required to establish a Dirichlet boundary value problem for the metric. Here $\ga$ denotes the determinant of the induced metric at the boundary $\partial \mathcal{M}$ and $K$ is the trace over extrinsic curvature.

The novel Born--Infeld-like boundary term $I_{\textrm{\tiny BI}}$ and its implications for thermodynamics will be derived and discussed in detail in the body of our paper. The functions $Q$, $w$ and $h$ appearing in it are constructed out of integrals of the functions $U$, $V$ and $f$:
\begin{align}
 Q(X) &= \int^X \!\!\!\!\!\extd y\, U(y)\\
 w(X) &= -2\int^X \!\!\!\!\!\extd y\, e^{Q(y)}V(y)\\
 h(X) &= \int^X \!\!\!\!\!\extd y\, \frac{e^{Q(y)}}{f(y)}
\end{align}
Their somewhat bizarre form has a natural explanation within 2d tensor-vector-scalar theories (or Einstein--Maxwell-dilaton-theories), which we shall review. Moreover, these functions have to obey certain inequalities, which we shall derive. The inequalities imply, among other things, that the electrostatic potential is confining.

Another key result is that we recover standard thermodynamics with the usual first law, say, for free enthalpy:
\eq{
\extd G(T,\,p,\,\dots) = -S\extd T + V\extd p + \textrm{other\;work terms}
}{eq:intro3}
Here $S$ is the 2d dilaton gravity analogue of the Bekenstein--Hawking entropy, $T$ is essentially the Hawking temperature, $-p$ is proportional to the cosmological constant and $V$ is the `volume' of the black hole -- a quantity that we will explain in detail. Thus, standard thermodynamics can be applied to 2d dilaton gravity models with a state-dependent cosmological constant.

The outline of this work is as follows: in order to make this paper self-contained we review some of the main results of \cite{Grumiller:2007ju} in section \ref{se:2} and generalize them to confining electrostatic potentials; in section \ref{se:3} we demonstrate how to treat the cosmological constant as a thermodynamical variable in a natural way in 2d dilaton gravity and derive the main results; in section \ref{se:4} we discuss general implications for thermodynamics and show the equivalence of thermodynamic and geometric volume definitions; in section \ref{se:7} we address further developments: we present alternative confinement conditions, unravel a Schottky anomaly in the specific heat and treat the microcanonical ensemble; in section \ref{se:5} we apply our results to specific examples, namely AdS-Schwarzschild--Tangherlini and BTZ black holes, as well as the Jackiw--Teitelboim model.

In this paper we use the conventions of \cite{Grumiller:2007ju}, setting the 2d Newton constant to $8\pi G_2=1$ and working exclusively in Euclidean signature.

\section{Thermodynamics review}\label{se:2}

In the following we review some of the main results of \cite{Grumiller:2007ju}, starting with the (holographically un-renormalized) Euclidean action of 2d dilaton gravity,
\begin{equation}
I = I_{\textrm{\tiny bulk}} + I_{\textrm{\tiny GHY}}
\label{bulkaction}
\end{equation}
with
\begin{align}
  I_{\textrm{\tiny bulk}} &= -\tfrac{1}{2}\int_{\mathcal{M}}\!\!\!\extd^2x \sqrt{g}\,\left(XR-U(\nabla X)^2-2V\right) \label{eq:bulk}\\
  I_{\textrm{\tiny GHY}} &= -\int_{\partial \mathcal{M}}\!\!\!\!\!\extd x\sqrt{\gamma}\, XK\,, 
\end{align}
where $X$ is the dilaton field, $R$ the Ricci scalar, $U$ is the kinetic potential and $V$ the dilaton self-interaction potential, both of which are model dependent functions of the dilaton $X$. The boundary term is analogous to the Gibbons--Hawking--York boundary term, with $\gamma$ the induced metric on $\partial \mathcal{M}$ and $K$ the trace of the extrinsic curvature. 

\subsection{Classical solutions}

The equations of motion (EOM) for the action \eqref{eq:bulk} are given by
\begin{align}
 \nabla_\mu \partial_\nu X-g_{\mu \nu}\nabla^2X-g_{\mu \nu} V \qquad\quad & \nonumber \\
 + U(\partial_{\mu}X)(\partial_{\nu}X)-\tfrac{1}{2}g_{\mu\nu}U\,(\partial X)^2 &=0 \label{EOM1} \\
 R+U^\prime(\partial X)^2+2U\nabla^2X-2V^\prime&=0 \label{EOM2}
\end{align}
where prime denotes a derivative with respect to the dilaton $X$. Taking the trace of the first EOM \eqref{EOM1} yields the useful relation $\nabla^2 X=-2V$. 

The space of all classical solutions to the EOM \eqref{EOM1}, \eqref{EOM2} falls into two disjoint classes: 1.~constant dilaton vacua, with $X=X_0=\rm const.$, which exist only for special models and require infinite fine-tuning of the value of the dilaton field such that $V(X_0)=0$; in that case the metric is maximally symmetric, i.e., either Minkowski, AdS or de~Sitter, depending on the sign of the Ricci scalar $R=2V^\prime(X_0)$; and 2.~linear dilaton vacua, which exist generically as solutions to arbitrary dilaton gravity models and have dilaton fields that are not constant. In this paper we exclusively focus on the latter.

Locally, all linear dilaton solutions are parametrized by a single constant of motion $M$ and take the form
\begin{equation}
X=X(r) \qquad \extd s^2=\xi(r)\extd\tau^2+\frac{1}{\xi(r)}\,\extd r^2 \label{solution}
\end{equation}
with
\begin{align}
\partial_rX&=e^{-Q(X)} \\
\xi(X)&=w(X)e^{Q(X)}\Big(1-\frac{2M}{w(X)}\Big) \label{killnorm} \,.
\end{align}
The functions $Q(X)$ and $w(X)$ are defined as \footnote{%
While we shall not need it in the present work, for the reader who is not familiar with 2d dilaton gravity we mention another interesting property of the functions $w(X)$ and $Q(X)$.
The function $w(X)$ is invariant under dilaton-dependent Weyl rescalings, $g_{\mu\nu}\to \exp{[2\Omega(X)]} \, g_{\mu\nu}$. By contrast, the function $Q(X)$ is shifted linearly with the conformal factor $2\Omega(X)$. Thus, physical quantities that depend on the potentials $U(X)$ and $V(X)$ solely through the combination $w(X)$ (or derivatives thereof) are Weyl invariant, while quantities that depend also on $Q(X)$ (or derivatives thereof) are not Weyl invariant. 
}
\begin{align}
Q(X)&:= Q_0+\int^{X}\!\!\!\!\!\extd y\,U(y) \label{eq:Q} \\
w(X)&:= w_0-2\int^{X}\!\!\!\!\!\extd y\,V(y)e^{Q(y)} \label{eq:w} \,.
\end{align}
By an appropriate choice of $w_0$ the constant of motion $M$ can be restricted to a convenient range, for instance non-negative values, $M\ge0$. This then fixes the ambiguity from the integration constant in the definition of the function $w$. The ambiguity in the definition of the function $Q$ is fixed by an appropriate choice of $Q_0$, which in turn can be absorbed into a rescaling of the coordinates. We shall henceforth assume that a suitable zero point has been chosen for the mass $M$ and suitable physical units for the coordinates, so that $w_0$ and $Q_0$ are fixed (for example, to zero).

All solutions \eqref{solution} exhibit a Killing vector field $\partial_\tau$ with norm $\xi(X)$ that vanishes at the `Killing horizon' $X=X_h$. To be precise, we mean that in the Minkowskian version of \eqref{solution} constant $X$ hypersurfaces that obey $\xi(X_h)=0$ are Killing horizons. In the corresponding Euclidean space the `Killing horizon' reduces to a point that provides either a lower or an upper bound on the value of the dilaton field. The `ground state Killing norm' $\xi(X)|_{M=0}$ will be denoted as $\xi_0$.  

In the models that we consider there is typically at least one Killing horizon. The dilaton field $X$ is then restricted to the positive, semi-infinite interval
\begin{equation}
X_h\le X < \infty\,,
\end{equation}
where $X_h$ denotes the outermost Killing horizon, i.e., the Killing horizon at the largest possible value of $X$. 

As in \cite{Grumiller:2007ju} we assume $\lim_{X\to\infty} w(X)=+\infty$, which captures many models of interest, in particular black holes in asymptotically AdS spacetimes. Thus, the asymptotic behavior of the metric is given by $\xi_{0}(X)$, as is easily seen from \eqref{killnorm}. 

Before discussing black hole thermodynamics we conclude this part of our review with a standard Euclidean argument. In order to avoid conical defects at the horizon, the Euclidean time is assumed to be periodic $\tau \sim \tau + \beta$, with 
\begin{equation}
\beta=\frac{4\pi}{\partial_r\xi}\bigg|_{r_h}\,.
\label{eq:beta}
\end{equation}
The proper local temperature $T(X_c)$ evaluated at a locus $X=X_c$ is equal to $\beta^{-1}$ times a redshift (`Tolman') factor 
\begin{equation}
T(X_c)=\frac{1}{\sqrt{\xi(X_c)}}\,\beta^{-1}\,.
\label{eq:lalapetz}
\end{equation}
Thus, in models where $\xi(X_{c}) \to 1$ as $X_c \to \infty$ the inverse period $\beta^{-1}$ is the temperature $T$ as measured by an observer at infinity.


\subsection{Thermodynamics from Euclidean path integral}

In order to study the thermodynamics of this model, the Euclidean path integral with appropriate boundary conditions is expanded around the classical solutions of the EOM. In the semi-classical limit the dominant contributions are the classical solutions. Then the Euclidean partition function is given by
\begin{equation}
Z\sim \sum_{g_{cl},\,X_{cl}}\exp{\left(-I\left[g_{cl},\,X_{cl}\right]\right)} \times Z_{\textrm{\tiny Gauss}} \times Z_{\textrm{\tiny ho}}\label{pathfluc}
\end{equation}
with  
\eq{
Z_{\textrm{\tiny Gauss}}=\int \mathcal{D}\delta g\mathcal{D}\delta X \exp{\left(-\tfrac{1}{2}\delta^2I\left[g_{cl},X_{cl};\delta g, \delta X\right]\right)}
}{eq:angelinajolie}
and higher order corrections contained in $Z_{\textrm{\tiny ho}}$. The sum in \eqref{pathfluc} extends over all classical solutions $g_{cl}$, $X_{cl}$ from \eqref{solution} compatible with the boundary conditions that we impose in order to evaluate the path integral. In our case, these boundary conditions will always specify the locus of a cut-off surface $X=X_c$ and the local temperature \eqref{eq:lalapetz} at that surface. In many cases it makes sense to think of these boundary conditions as coupling a finite gravitational system to a thermal reservoir at proper local temperature $T(X_c)$. 

The free energy in the classical approximation is determined by the on-shell action through
\eq{
F =  -T\,\ln Z \approx T \,I\big[\hat g_{cl},\,\hat X_{cl}\big]
}{eq:F}
where $\hat g_{cl}$, $\hat X_{cl}$ denotes the dominant saddle-point in the sum \eqref{pathfluc}. Sub-dominant saddle-points are then non-perturbative (``instanton'') corrections, while the neglected Gaussian and higher order terms are perturbative corrections. In the present work we shall not be concerned with either of these corrections. 

The thermodynamical ensemble is determined by the choice of boundary conditions used in the evaluation of the path integral. We shall always keep fixed the proper temperature at the boundary, so that our free energy is either the Helmholtz or Gibbs free energy (or some generalization thereof). Once the free energy is known, all other thermodynamical quantities of interest are determined, either through partial derivatives or Legendre transformations of the free energy.

The semiclassical approximation is well defined only if 
\begin{itemize}
\item the on-shell action is bounded, and
\item the first variation of the action vanishes on-shell for all variations that preserve the boundary conditions.
\end{itemize}
As expected on general grounds (and shown explicitly in \cite{Grumiller:2007ju}) the action (\ref{bulkaction}) has neither of these properties for the boundary conditions we are interested in. These shortcomings are remedied by adding a (holographic \mbox{counter-)}\-term $I_{\textrm{\tiny CT}}$ to (\ref{bulkaction}), yielding an improved action
\begin{equation}
\Gamma= I_{\textrm{\tiny bulk}}+I_{\textrm{\tiny GHY}}+I_{\textrm{\tiny CT}} \label{fullaction}
\end{equation}
with 
\begin{equation}
I_{\textrm{\tiny CT}} =\int_{\partial \mathcal{M}}\!\!\!\!\!\extd x\sqrt{\gamma}\,\sqrt{e^{-Q(X)}w(X)}\,.\label{oldct}
\end{equation} 
It was later shown that the same boundary term can be derived from requiring supersymmetry in the presence of boundaries \cite{Grumiller:2009dx}, \footnote{Technically, the key observation here is that the counterterm Lagrangian $\sqrt{e^{-Q(X)}w(X)}=|u(X)|$ is given by the pre-potential $u(X)$, which is the lowest component of the pre-potential supergravity  multiplet.}.

In 2d dilaton gravity every sufficiently regular function $J(X)$ of the dilaton field can be used to construct a conserved charge. This is so, because a current $j^\mu=\epsilon^{\mu\nu}\partial_\nu J$ is trivially divergence-free, $\partial_\mu j^\mu=0$. Hence 2d dilaton gravity exhibits an infinite number of completely equivalent conserved charges. Therefore, we may choose a suitable dilaton charge, which we denote by $D_c$, and demand that its value at the boundary be fixed. In the following we simply set
\begin{equation}
D_c=X_c\,.
\end{equation}
The heat bath surrounding the cavity fixes the local inverse temperature $\beta_{c}$ and dilaton charge $D_c$ as boundary conditions for the path integral. \par
The nonnegative solutions of
\begin{equation}
\beta_c=\sqrt{\xi(D_c,M)}\beta(M)\,,
\end{equation}
denote the possible values of the mass $M$ and thus the classical solutions consistent with the boundary conditions. The Helmholtz free energy $F$ may be obtained from the action via [see \eqref{eq:F}]
\begin{equation}
F_c(T_c,\,D_c)=T_c\,\Gamma(T_c,\,D_c)\,.
\end{equation}
With the above definitions this yields
\begin{equation}
F_c(T_c,\,D_c)=-2\pi X_h T_c+e^{-Q_c}\left(\sqrt{\xi_0}-\sqrt{\xi_c}\right)\,,
\label{eq:Fc}
\end{equation}
where a subscript $c$ denotes evaluation of the dilaton at the cut-off surface $X=X_c$. 

It is natural to consider the $X_c\to\infty$ limit of \eqref{eq:Fc} and other thermodynamic quantities, to obtain results for the full, non-compact spacetime. This limit is in general well-defined for the classical theory. However, in many theories the perturbative corrections \eqref{eq:angelinajolie} diverge as $X_c \to \infty$, and in those cases the cut-off cannot be removed. From the thermodynamic point of view this is so because the specific heat of the system becomes negative above some maximum value of $X_c$. In other words, the density of states grows too rapidly as a function of energy, and the canonical ensemble no longer exists.

Further details of 2d dilaton gravity thermodynamics, elaborations, applications and many examples can be found in \cite{Grumiller:2007ju}. Some earlier work on 2d dilaton gravity thermodynamics is \cite{Gegenberg:1994pv,Kunstatter:1998my,Davis:2004xi}.

\subsection{Charged black holes}\label{se:2.3}

The above results are easily generalized to charged black holes by adding a Maxwell term
\begin{equation}
I_{\textrm{\tiny Max}}=\int_\mathcal{M}\!\!\!\extd^2x\sqrt{g}\,f(X)F_{\mu \nu}F^{\mu \nu} \label{Maxwellaction}
\end{equation}
to the action (\ref{fullaction}). The function $f(X)$ describes the coupling of the abelian field strength $F_{\mu \nu}=\partial_\mu A_\nu-\partial_\nu A_\mu$ to the dilaton field. 
The EOM (Maxwell's equations) are solved by 
\begin{equation}
F_{\mu \nu}=\frac{q}{4f(X)} \,\varepsilon_{\mu \nu}\,,
\label{eq:Fmn}
\end{equation}
where $q$ denotes the electric charge, $\varepsilon_{\mu \nu}$ is the 2d epsilon-tensor with $\varepsilon_{\tau r}=+1$, and the factor $\frac{1}{4}$ is chosen for convenience. We will assume, for now, that all the fields fall off sufficiently rapidly so that no additional boundary terms are needed in the action.

The EOM for the action (\ref{fullaction}) with an additional Maxwell field are still solved by equation (\ref{solution}), but the Killing norm is changed to
\begin{equation}
\xi(X)=e^{Q(X)}\left(w(X)-2M+\tfrac{1}{4}q^2h(X)\right)\,. \label{chargedKill}
\end{equation}
The function $h(X)$ is defined as \footnote{%
Again there is an ambiguity in the definition of the function $h(X)\to h(X)+h_0$ from an integration constant $h_0$, and again it can be fixed by a suitable choice. By inspection of \eqref{chargedKill} constant shifts of the function $h(X)$ merely amount to a charge dependent redefinition of the mass parameter $M$. To reduce clutter we set $h_0=0$.
}
\begin{equation}
h(X)=\int^{X}\!\!\!\extd y\,\frac{e^{Q(y)}}{f(y)}\,. \label{eq:hdef}
\end{equation}
In axial gauge, $A_r=0$, the gauge potential is given by
\begin{equation}
A_\tau(X)=-\frac{q}{4}\left(h(X)-h(X_h)\right)+A_\tau(X_h)\,. 
\label{eq:A}
\end{equation}
Thus, the proper electrostatic potential relative to the horizon is 
\begin{equation}
\Phi(X)=\frac{A_\tau(X)-A_\tau(X_h)}{\sqrt{\xi(X)}}\,.
\end{equation}
Dirichlet boundary conditions on $A_{\tau}$ imply that this quantity is held fixed at the cut-off surface $X_c$.

With these results the Euclidean partition function \eqref{pathfluc} may be calculated in the classical approximation. 
For the boundary conditions that we impose the relevant thermodynamic potential is the Legendre transform of the Helmholtz free energy $F_c(T_c,\,D_c,\,q)$ with respect to the proper electrostatic potential $\Phi_c$. Since the charge $q$ and potential $\Phi_{c}$ play a role similar to a generic conserved charge $N$ with chemical potential $\mu$, we simply denote this potential by $Y_c= T_c\,\Gamma(T_c,\,D_c,\,\Phi_c)$.
\eq{
Y_c(T_c,\,D_c,\,\Phi_c) = F_c(T_c,\,D_c,\,q) - q\,\Phi_c
}{eq:Y}
The inverse Legendre transformation then leads to the Helmholtz free energy $F_c = Y_c + q\,\Phi_c$.
\begin{equation}
F_c(T_c,\,D_c,\,q)=-2\pi X_h T_c+e^{-Q_c}\left(\sqrt{\xi_0}-\sqrt{\xi_c}\,\right) \label{Freecharge}
\end{equation}
Evidently, this result is equal to the previous one \eqref{eq:Fc}.
The dependence on $q$ remains implicit in the Killing norm $\xi(X)$ and the locus of the horizon $X_h$, but does not show up explicitly. 

Notice that one may add an arbitrary number of Maxwell fields (in fact, even non-Abelian gauge fields) in this way. Each of them will introduce new conserved charges $q_i$ that capture some of the state-dependent information. We shall make use of this fact in the calculation of the first law for BTZ black holes. 

If we relax the assumption that the fields fall off sufficiently rapidly, then a new boundary term must be added to the action along with \eqref{Maxwellaction}. For a theory with Dirichlet boundary conditions on $A_{\mu}$, diffeomorphism invariance restricts the form of this boundary term to be
\begin{equation}
I_{\textrm{\tiny MCT}}=\int_{\partial\mathcal{M}}\!\!\!\!\!\extd x\sqrt{\ga}\,\mathcal{L}_{\textrm{\tiny MCT}}(A^{\mu}A_{\mu},\,X)\,.
\end{equation}
A simple example that requires such a boundary term can be found in \cite{Castro:2008ms}. In section \ref{se:3} we will consider theories with Neumann boundary conditions on $A_{\mu}$, and hence we will obtain a different sort of boundary term.

\subsection{Confinement in 2d dilaton gravity}\label{se:confinement}

In the discussion of charged black holes above we have assumed that the gauge potential decays asymptotically sufficiently fast so that no modification of the boundary term \eqref{oldct} and of the free energy \eqref{Freecharge} arises. However, this is not necessarily true, and in particular it is not true for confining potentials. Before addressing how to improve the boundary term we address the case of confining potentials.

To this end we consider now specific classes of 2d Einstein--Maxwell-dilaton theories that obey the inequalities
\begin{align}
 & \lim_{X\to\infty} w(X) \to +\infty \label{eq:con1} \\
 & \lim_{X\to\infty} |f(X)V(X)| < \infty \,. \label{eq:con2}
\end{align}
We show now that these inequalities imply confinement of the $U(1)$ charge in the sense that the gauge potential $A_\mu(X)$ diverges  to $\pm\infty$ for large $X$. Before we explain why the attribute `confinement' is justified we check that our claim is technically correct.

The result \eqref{eq:A} together with the definition \eqref{eq:w} imply
\eq{
A_\tau(X\to\infty) = \frac q4\, \int^{X}\!\!\!\extd y\,\frac{w^\prime(y)}{f(y)V(y)} \,. 
}{eq:Acon}
Taking absolute values and exploiting the inequality \eqref{eq:con2} yields
\eq{
\big| A_\tau(X\to\infty) \big| > N^2\, w(X\to\infty) \,, 
}{eq:Acon2}
where $N$ is some non-vanishing real number. The inequality \eqref{eq:Acon2} together with the assumption \eqref{eq:con1} prove our claim that the gauge potential $A_\mu(X)$ diverges to $\pm\infty$ for large $X$.

Of course, one could just define a confining potential in 2d Einstein--Maxwell-dilaton gravity by the properties \eqref{eq:con1}, \eqref{eq:con2}. However, it is useful to clarify why it is justified to call a gauge potential that diverges at large $X$ `confining'. Physically, the decisive property of a confining potential is that it takes infinite energy to separate two charges. If we take as one of the charges the charged black hole spacetime and as the other some test-charge, then a measure for the electrostatic energy is the product of the proper electrostatic potential times the charge $\hat q$ of the test-particle, measured for instance in units of the proper temperature (in order to cancel redshift factors, assuming some non-extremal configuration with $T\neq 0$). Thus, if the ratio
\eq{
\frac{\hat q \, \Phi_c}{T_c} = \hat q\, \frac{A_\tau(X_c)}{T} + {\cal O}(1)
}{eq:con3}
stays finite in the limit $X_c\to\infty$ then the gauge potential is not confining. On the other hand, if the expression \eqref{eq:con3} tends to $\pm\infty$ in the limit $X_c\to\infty$ then the gauge potential is confining. From the right hand side of \eqref{eq:con3} we therefore see that the gauge potential is confining precisely if $A_\tau(X)$ diverges to $\pm\infty$ for large $X$.

A simple class of examples for confining potentials is provided by minimally coupled Maxwell fields, $f(X)=\rm const.$, and vanishing kinetic potential $U(X)=0$. In that case the gauge potential \eqref{eq:Acon} integrates to a function that is linear in $X\propto r$. We recover the well-known fact that the 2d electrostatic potential grows linearly in $r$ and is thus confining. 

A simple class of examples for non-confining potentials are spherically reduced models with $f(X)\propto X$,
\eq{
 U(X)=-\frac{d-2}{(d-1)X}\quad \textrm{and}\quad V(X)\propto X^{1-2/(d-1)}\,, 
}{eq:UVsrg}
which violate the inequality \eqref{eq:con2} and reproduce the correct $d$-dimensional Coulomb law $\Phi\propto 1/r^{d-2}$ in $d>2$ spatial dimensions.

In the next section we stick to a single confining Maxwell field with a specific type of coupling function $f$ in order to describe a state dependent cosmological constant $\Lambda$.

\section{State dependent $\Lambda$}\label{se:3}

In this section we implement a state dependent cosmological constant $\Lambda$ within 2d dilaton gravity. As a first step we clarify how a state \emph{in}dependent cosmological constant appears in 2d dilaton gravity. 

\subsection{Defining a 2d cosmological constant}

The notion of a cosmological constant is ambiguous in 2d dilaton gravity. In any dimension greater than two it implies simultaneously two things: 1.~$\Lambda$ is a parameter in the action that is multiplied by the volume form, and 2.~Vacuum solutions asymptote to constant curvature spaces, i.e., de~Sitter or AdS, depending on the sign of $\Lambda$. In 2d dilaton gravity, however, adding a constant to the potential $V$ does not lead to asymptotically constant curvature spaces in general. Instead, in order to obtain such spaces one has to add a linear term in the dilaton to the potential $V$. This is seen most easily from the EOM \eqref{EOM2} for kinetic potentials $U$ that vanish asymptotically sufficiently fast: if $V$ is constant then the Ricci scalar vanishes asymptotically. However, if $V=\Lambda X$ then the Ricci scalar asymptotes to $R=2\Lambda$.

In our work we always use the second notion of `cosmological constant', i.e., when we have positive (negative) $\Lambda$ and the cosmological constant term in the dilaton potential 
\eq{
V=\Lambda X + V_{\textrm{\tiny rest}} 
}{eq:V}
dominates at large values of $X$ then the metric asymptotes to 2d de~Sitter or AdS space. 

In fact, there is a natural interpretation of $V=\Lambda X$ in the context of theories that emerge from dimensional reduction of sufficiently symmetric higher-dimensional theories of gravity. Namely, the original volume form is proportional to the 2d volume form times the dilaton field times some (irrelevant) constant volume factor, the value of which depends on the internal space. We recall this now explicitly for spherical reduction.

The bulk term in the Einstein--Hilbert action for pure Einstein gravity with a cosmological constant in $d+1$ dimensions is
\begin{equation}
I_{d+1}=-\frac{1}{16\pi G_{d+1}}\int_{\mathcal{M}}\!\!\!\extd^{d+1}x\sqrt{g_{d+1}}\,\left(R_{d+1}-2\Lambda\right)\,. \label{EHhigh}
\end{equation}
Spherical symmetry implies that the $(d+1)$-dimensional metric can be brought into the adapted form \cite{Berger:1972pg,Unruh:1976db,Benguria:1977in,Thomi:1984na,Hajicek:1984mz}
\begin{equation}
\extd s^2=g_{\mu \nu}\,\extd x^{\mu}\extd x^{\nu}+(G_{d+1})^{\frac{2}{d-1}}\varphi^2(x^\mu)\,\extd\Omega^2_{S^{d-1}}
\end{equation}
where $\mu, \,\nu\in\{0,\,1\}$, $\varphi(x^\mu)$ is essentially the surface radius and $\extd\Omega^2_{S^{d-1}}$ is the line-element of the round $(d-1)$-sphere.
Inserting this Ansatz into the Einstein Hilbert action (\ref{EHhigh}) upon integrating out the $(d-1)$-sphere yields
\begin{equation}
I_{\textrm{\tiny SRG}} = -\frac{A_{d-1}}{16\pi}\,\int_{\mathcal{M}}\!\!\!\extd^2x \sqrt{g}\, \varphi(r)^{d-1}\left(R_{d+1}-2\Lambda \right)\,, \label{twomet}
\end{equation}
where $A_{d-1}$ denotes the solid angle subtended by a $(d-1)$-sphere. 
The action \eqref{twomet} is now a special case of the 2d dilaton gravity bulk action \eqref{eq:bulk} if we express the $(d+1)$-dimensional Ricci scalar in terms of 2d quantities (see for instance Eq.~(C.10) in \cite{Grumiller:2001ea}) 
\eq{
R_{d+1} = R + \frac{(d-1)(d-2)}{\varphi^2}\,\big(G^{\frac{2}{1-d}}_{d+1}- (\nabla\varphi)^2\big) - \frac{2(d-1)}{\varphi}\,\nabla^2\varphi
}{eq:Rd}
and define the dilaton field $X(r)$ as
\begin{equation}
X(r)=\frac{A_{d-1}}{8\pi G_{d+1}}G_{d+1}\varphi(r)^{d-1}\,.
\label{eq:dil}
\end{equation} 
The result (\ref{twomet}) makes it explicit that the dilaton field couples to the Ricci scalar and the cosmological constant in the same way, i.e., linearly in terms of $X$ as defined in \eqref{eq:dil}. We recover exactly the potentials \eqref{eq:UVsrg}, but with $V$ replaced by $V+\Lambda\, X$, concurrent with \eqref{eq:V}.

\subsection{Converting $\Lambda$ into a $U(1)$ charge}

Having agreed that the term we want to add to the dilaton potential $V$ is given by $\Lambda X$ we proceed now with one of our main goals, namely to make $\Lambda$ a state dependent quantity rather than a parameter in the action. In fact, there is a well-known procedure in 2d dilaton gravity to convert parameters in the action into constants of motion by `integrating in' Maxwell fields (see appendix B of \cite{Grumiller:2005sq} for a summary).

Translating the general procedure to the present context we add a Maxwell term (\ref{Maxwellaction}) with the specific coupling
\eq{
f(X)=\frac{1}{X}\,.
}{eq:important} 
This choice leads to the non-minimally coupled Maxwell action
\eq{
I_{\textrm{\tiny Max}} = \int_\mathcal{M}\!\!\!\extd^2x\sqrt{g}\,\frac{1}{X}\,F_{\mu\nu} F^{\mu\nu}\,.
}{eq:important2}
Inserting the on-shell value \eqref{eq:Fmn} for the field strength establishes 
\begin{equation}
I_{\textrm{\tiny Max}}\big|_{\textrm{\tiny EOM}} = \int_\mathcal{M}\!\!\!\extd^2x\sqrt{g}\,X\,\frac{q^2}{8}\,.
\end{equation}
From the above it seems reasonable to set $q^2\propto \Lambda$. The correct relation (in particular, the correct sign and factor) between $U(1)$ charge and cosmological constant,
\begin{equation}
\frac{q^2}{8}=-\Lambda\,,\label{eq:ccq}
\end{equation}
can be deduced from comparing the full bulk action $I_{\textrm{\tiny bulk}}$ \eqref{eq:intro1.5}, with the $U(1)$ field integrated out, with \eqref{twomet}. Namely, if we disregard for a moment the terms containing the potentials $U$ and $V$ we obtain
\begin{align}
 I_{\textrm{\tiny bulk}}\big|_{U=V=0} &= -\tfrac{1}{2}\int_{\mathcal{M}}\!\!\!\extd^2x \sqrt{g} \, \big(XR  + f(X)F^{\mu \nu}F_{\mu\nu}\big) \nonumber\\ & \quad -4\int_{\mathcal{M}}\!\!\!\extd^2x \sqrt{g}\,\partial_{r}\big(f(X)F^{r \tau}A_{\tau}\big)
\end{align}
which on-shell reduces to
\begin{equation}
I_{\textrm{\tiny bulk}}\big|_{U=V=0}= -\tfrac{1}{2}\int_{\mathcal{M}}\!\!\!\extd^2x \sqrt{g} \, \Big(XR+\frac{q^2}{4}\frac{1}{f(X)}\Big)\,.
\end{equation}
The result above with the identification \eqref{eq:important} then leads to the equality \eqref{eq:ccq} upon comparison with \eqref{twomet}-\eqref{eq:dil}.

Thus, we have succeeded in converting the cosmological constant into a state dependent parameter, namely a conserved $U(1)$ charge. For real charges $q$ the cosmological constant is negative, so henceforth we restrict ourselves to discussions of asymptotically AdS spacetimes.

\subsection{Confinement of $\Lambda$}\label{se:3.3}

The cosmological constant emerges as a thermodynamic variable naturally this way, since we may regard it as the charge of a Maxwell field with the peculiar coupling \eqref{eq:important}. Let us now check under which conditions this leads to a confining potential. The inequality \eqref{eq:con2} holds provided the dilaton potential $V(X)$ obeys 
\eq{
\lim_{X\to\infty} \Big|\frac{V(X)}{X}\Big| < \infty   \,.
}{eq:Vineq}
In other words, as long as the original dilaton potential does not grow faster than $X$ at large values of the dilaton, the confinement inequality \eqref{eq:con2} holds. To ensure this inequality we assume from now on that $V(X)$ grows slower than $X$  at large values of the dilaton,
\eq{
V(X\to\infty) \propto X^\al\qquad \al < 1\,.
}{eq:Vass}
This assumption holds for all values of dimension $2<d<\infty$ for spherically reduced models, see \eqref{eq:UVsrg} \footnote{Saturation of the inequality, $\alpha=1$, is compatible with the confinement condition \eqref{eq:Vineq} as well, but then $V$ scales in the same way as the cosmological constant. The value $\alpha=1$ is obtained, e.g., in the $d\rightarrow \infty$ Schwarzschild--Tangherlini case \cite{Emparan:2013xia}, for the Witten black hole \cite{Mandal:1991tz,Elitzur:1991cb,Witten:1991yr} and the model by Callan, Giddings, Harvey and Strominger \cite{Callan:1992rs}. We do not consider these cases in the present work, because they provide an exception to the rule that potentials linear in $X$ lead to asymptotically (A)dS spacetimes. This is so, because the terms containing the kinetic potential $U$ in the expression for curvature \eqref{EOM2} cancel the contribution from the dilaton self-interaction so that $R\to 0$.}.

A key property of confining charges in general and the charge describing an effective cosmological constant in particular is that the asymptotic behavior of the Killing norm in the metric \eqref{solution} is dominated by the charge term
\eq{
\xi = -2\Lambda\,e^{Q(X)}\,h(X) + \dots 
}{eq:xiasy}
The fact that all other terms in the Killing norm are subleading follows from the inequalities \eqref{eq:con1} and \eqref{eq:con2}.

The discussion in section \ref{se:2.3}, in particular the use of the boundary term \eqref{oldct}, assumed that the term in the metric highlighted in \eqref{eq:xiasy} was asymptotically subleading. This is no longer the case, so we need to find a suitable boundary term that gives a 
well-defined variational principle. We have summarized this main result, which is a natural generalization of \eqref{oldct}, in equations \eqref{eq:intro1}-\eqref{eq:intro2} of the introduction. There are two main differences to the old results used in section \ref{se:2}. 

Firstly, the action now includes a bulk total derivative term
\eq{
-4\int_{\mathcal{M}}\!\!\!\extd^2x \sqrt{g}\,\nabla_{\mu}(fF^{\mu \nu}A_{\nu}).
}{eq:totder}
This term arises from requiring that the on-shell action is a function of $q$ rather than $\Phi_c$.
In fact, the term \eqref{eq:totder} corresponds on-shell to $q\,\Phi_c$, the expression appearing in the (inverse) Legendre transformation of \eqref{eq:Y}. So this term ensures that we are in a thermodynamic ensemble where the cosmological constant $\Lambda$ is fixed as part of the boundary conditions, as we shall discuss in detail in the next section.
Note that a corresponding term was already considered in \cite{Brown:1988kg} in a similar model. 

Secondly, the boundary term \eqref{oldct} is extended to a Born--Infeld type of boundary action.
\eq{
I_{\textrm{\tiny CT}}=\!\int_{\partial \mathcal{M}}\!\!\!\!\!\extd x \sqrt{\gamma}\, \sqrt{e^{-Q(X)} \!\left(w(X)+2 F_{\mu \nu}F^{\mu \nu}f(X)^2h(X)\right)} 
}{eq:newct}
This is one of our main results. We explain now why (and under which conditions) this result is correct, starting first with an on-shell argument, showing next that the result above leads to a meaningful expression for the free energy and arguing finally that the full action  with the boundary term \eqref{eq:newct} leads to a well-defined variational principle, in the sense that the first variation of the action \eqref{eq:intro1} vanishes for all variations that preserve our boundary conditions.

With hindsight, the Born--Infeld boundary term \eqref{eq:newct} is a natural generalization of (\ref{oldct}) that reduces on-shell to the ground state Killing norm.
\eq{
\xi_0(X)=e^{Q(X)}\left(w(X)-2\Lambda\, h(X)\right)
}{eq:gk}
In order to ensure that the ground state Killing norm is also the asymptotic Killing norm
\eq{
\xi(X)=\xi_0(X)-2e^{Q(X)} M = \xi_0(X) + \textrm{subleading}
}{eq:ak}
we have to impose a restriction on the kinetic potential $U(X)$ or, equivalently, on the function $h(X)$, viz.
\eq{
\lim_{X\to\infty}|h(X)|\to\infty\,.
}{eq:hres}
Expressed as a condition on the function $Q(X)$ the condition \eqref{eq:hres} yields the inequality
\eq{
\lim_{X\to\infty} e^{Q(X)} > \frac{N^2}{X^2}
}{eq:Qres}
where $N$ is some non-vanishing real number. Translating this into a condition on the kinetic potential $U(X)$ finally establishes~\footnote{%
For the reader not familiar with older literature on 2d dilaton gravity we mention that the limiting case $U=-2/X$ is special as far as the causal structure of space-time is concerned and divides the space of models concerning the asymptotic and singularity structures into two regions, one of which obeys the inequality \eqref{eq:Ures}. See for instance Fig.~3.12 in \cite{Grumiller:2002nm} where this critical line corresponds to $a=2$ in their conventions.
}
\eq{
\lim_{X\to\infty} U(X) > -\frac{2}{X}\,.
}{eq:Ures}
In the present work we are going to consider exclusively models that obey the inequality \eqref{eq:Ures}. Note that this includes the special case of vanishing kinetic potential, $U=0$, as well as spherically reduced gravity \eqref{eq:UVsrg} for spatial dimensions $1<d\leq \infty$.

Following the steps reviewed in section \ref{se:2}, the on-shell action \eqref{eq:intro1} yields the same expression for the thermodynamic potential as equation \eqref{Freecharge}, with $\xi_0$ now defined as above. But since we wish to interpret the cosmological constant $\Lambda$ as (negative) pressure, the thermodynamic potential for this ensemble should now be regarded as the Gibbs free energy, rather than the Helmholtz free energy.
\begin{equation}
G_c(T_c,\,\Lambda,\,D_c)=-2\pi X_h T_c+e^{-Q_c}\left(\sqrt{\xi_0}-\sqrt{\xi_c}\right) \label{freenew}
\end{equation}
This point will be discussed in more detail in the beginning of the next section.

The new action (\ref{eq:intro1}) yields a well-defined variational principle provided the dilaton field obeys Dirichlet boundary conditions 
 \eq{
 \de\ln X\big|_{\partial{\mathcal M}} = 0
 }{eq:Xbc}
and the asymptotic behavior of the variation of the Killing norm $\de \xi$ obeys 
\eq{
\delta \xi=U\xi\,\delta X-2e^Q h \,\delta \Lambda + \dots  
}{constraint}
where the ellipsis denotes asymptotically subleading terms, given our assumptions on the potentials~\footnote{The conditions \eqref{eq:Xbc} and \eqref{constraint} ensure that the variation of the action vanishes when the EOM are satisfied. But the EOM may also result in kinematic constraints on the asymptotic form of the fields. These constraints would imply additional conditions on the field variations.}. 

The above discussion was concerned with treating $\Lambda$ as a particular example of a confining $U(1)$ charge, but as mentioned in section \ref{se:confinement}, the action \eqref{eq:intro1} provides a well-defined variational principle and the correct thermodynamics for generic confining $U(1)$ fields subject to the conditions \eqref{eq:con1} and \eqref{eq:con2}, provided we are in the thermodynamic ensemble where $q$ is kept fixed. For related work in higher dimensions, see  \cite{Gonzalez:2009nn}.

Without the the bulk total derivative term \eqref{eq:totder} the action \eqref{eq:intro1} diverges when evaluated on-shell. In the case of AdS this is an unwanted property, which motivated the addition of the term \eqref{eq:totder} to the full action. However, for generic confining potentials the divergence of the on-shell action has a physical interpretation, namely the infinite amount of energy required to bring a charge out to infinity, and is therefore a feature of the on-shell action appropriate for the ensemble in which the electrostatic potential $\Phi_c$ is kept fixed. Thus, when studying generic confining potentials in that ensemble one removes the term \eqref{eq:totder} from the full action \eqref{eq:intro1} and keeps only the Born--Infeld boundary term \eqref{eq:newct}.

In the next section we exploit the results of the present section to discuss thermodynamics in the presence of a state-dependent cosmological constant $\Lambda<0$.

\section{${\Lambda}$-thermodynamics}\label{se:4}

In the following we are going to study black hole thermodynamics with the cosmological constant promoted to a thermodynamic variable. We start by clarifying some nomenclature regarding energy versus enthalpy.

In the previous sections the on-shell action determined the Legendre transform of the Helmholtz free energy with respect to the pair $q$ (the charge) and $\Phi_c$ (proper electrostatic potential) \eqref{eq:Y}. The Helmholtz free energy \eqref{Freecharge} depends on temperature $T_c$, the electric charge $q$, and the dilaton charge $X_c$. The label ``Helmholtz free energy'' was justified there, since the charge is usually an extensive quantity (doubling the volume doubles the charge), while the proper electrostatic potential $\Phi_c$ is an intensive quantity. However, when interpreting the (square of the) electric charge as cosmological constant through the identification \eqref{eq:ccq} we should not consider $\Lambda$ as an extensive quantity (doubling the volume should not change the cosmological constant). Instead, $\Lambda$ is now intensive and acts as a pressure and, as we shall demonstrate below, its conjugate variable has the properties expected of a `volume'. Therefore, the expression \eqref{freenew} is actually the Gibbs free energy, also known as free enthalpy.

\subsection{Entropy, chemical potential and volume}

The entropy is obtained from the free enthalpy $G_c$ in the usual way
\eq{
S = - \frac{\partial G_c}{\partial T_c}\Big|_{D_c,\,\Lambda} = 2\pi X_h\,. 
}{eq:S}
This result for entropy is the same as the ones that follow from \eqref{eq:Fc} and \eqref{Freecharge} \cite{Grumiller:2007ju}.
Thus, a variable cosmological constant [or more generally, a confining $U(1)$ charge] leaves the form of entropy unchanged and enters only in determining the locus of the horizon $X_h$. 

The entropy \eqref{eq:S} coincides with the Wald entropy \cite{Wald:1993nt} and, in the case of dimensionally reduced Einstein gravity, correctly captures the higher-dimensional Bekenstein--Hawking area law. In that sense, the relation \eqref{eq:S} between entropy and $X_h$ is the 2d analogue of the Bekenstein--Hawking law. For dilaton gravity models coming from spherical reduction, $X_h$ is proportional to the horizon area of the higher-dimensional theory \eqref{eq:dil}, which allows to write \eqref{eq:S} in the usual way
\begin{equation}
S=\frac{A_h}{4G_{d+1}}.
\end{equation}
A Bekenstein--Hawking law for intrinsically 2d models can be formulated as well, if one associates the horizon to one connected component of a sphere in one dimension (which consists of two disjoint points) $A_h=A_1/2=1$ and introduces the effective Newton coupling $G_{\textrm{\tiny eff}}=G_2/X$ for scalar-tensor theories, where $X$ should be evaluated at some scale. Since the horizon is the only scale available, one can define $G_{\textrm{\tiny eff}}=G_2/X_h$, which allows to rewrite \eqref{eq:S} in the suggestive form \cite{Grumiller:2007ju}
\begin{equation}
S=\frac{A_h}{4 G_{\textrm{\tiny eff}}}\,.
\end{equation}

The variable conjugate to the dilaton charge $D_c$ used to determine the location of the cavity around the black hole is called the dilaton chemical potential $\psi_c$, which is obtained from the free enthalpy by
\begin{multline}
\psi_c=-\frac{\partial G_c}{\partial D_c}\Big|_{T_c,\,\Lambda} =-\frac{1}{2}U_ce^{-Q_c}\big(\sqrt{\xi_c}-\sqrt{\xi_0}\big)+\\
\Big(\frac{1}{2}w^{\prime}_c-\Lambda h^{\prime}_c\Big)\Big(\frac{1}{\sqrt{\xi_c}}-\frac{1}{\sqrt{\xi_0}}\Big)\,. \label{SSpsi}
\end{multline}
In the case $\Lambda=0$ this expression reduces to the one found in \cite{Grumiller:2007ju}. The presence of a negative cosmological constant increases the value of the dilaton chemical potential at the cut-off surface.

The third pair of variables is given by $\Lambda$ and its conjugate $\Theta_c$. These correspond to the pair $q$ and $\Phi_c$ when the interpretation of the $U(1)$ charge as a cosmological constant is not invoked. The quantity $\Theta_c$, which we shall refer to as the `thermodynamical volume', is obtained from the Gibbs free energy by
\eq{
\Theta_c =-\frac{\partial G_c}{\partial \Lambda}\Big|_{T_c,\,D_c} =\frac{h_c}{\sqrt{\xi_0}}-\frac{h_c}{\sqrt{\xi_c}}+\frac{h_h}{\sqrt{\xi_c}}\,. 
}{Theta}
Here, the subscripts $h,c$ denote evaluation at the horizon or the cavity wall, respectively, and $h$ is the function defined in \eqref{eq:hdef}. The specific coupling $f(X)=1/X$, used to model $\Lambda$ as a $U(1)$ charge, yields
\begin{equation}
h(X)=\int^{X}\!\!\!\!\!\extd y\,ye^{Q(y)} \label{hcosmo}\,.
\end{equation} 
When evaluated at the horizon, this is precisely the definition for the geometric volume of a 2d black hole presented in \cite{Grumiller:2005zk} up to a constant factor. We shall elaborate on the connection between $\Theta_c$ and this quantity in section \ref{subsec:AsymptoticFirstLaw}.
%

\subsection{Enthalpy, energy and the first law}\label{se:4b}

With the above definitions of the variables of thermodynamic phase space, one can now formulate the first law of black hole thermodynamics. 
A Legendre transformation of the Gibbs free energy (\ref{freenew}) with respect to the pair $T_c,\, S$ yields the enthalpy
\begin{equation}
H_c(S,\,\Lambda,\,D_c)=e^{-Q_c}\big(\sqrt{\xi_0}-\sqrt{\xi_c}\,\big)\,.\label{internalnew}
\end{equation}
This expression for enthalpy coincides with the one obtained from the boundary stress tensor. 

The boundary stress tensor $T^{ab}$ is obtained from the usual Brown--York prescription.
\begin{equation}
T^{ab}= - \frac{2}{\sqrt{\gamma}}\frac{\delta \Gamma}{\delta \gamma_{ab}}\Big|_{\textrm{\tiny EOM}} \label{eq:BY}
\end{equation}
This yields
\begin{equation}
T^{\tau \tau}=-n^{\mu}\nabla_{\mu}X \gamma^{\tau \tau}+e^{-Q}\sqrt{\xi_0}\gamma^{\tau \tau}\, .
\end{equation}
In particular, the energy measured by an observer along the timelike unit Killing vector $u^{a}$ is
\begin{equation}
T^{ab}u_{a}u_{b}=e^{-Q_c}\big(\sqrt{\xi_0}-\sqrt{\xi_c}\big)\, ,
\end{equation}
which coincides with the enthalpy \eqref{internalnew}~\footnote{In the case of non-vanishing fields at the boundary other than the metric, the Brown--York prescription should be replaced by the definition for the boundary stress-tensor given in \cite{Hollands:2005ya}, which agrees with \eqref{eq:BY} if the field present is a scalar field. In the present case, the non-vanishing field at the boundary is the field-strength $F_{\mu \nu}$, which is dual to a scalar field in 2d. Therefore it is expected that both prescriptions give the same result, which is indeed the case.}.

The first law of thermodynamics can be deduced from $\eqref{internalnew}$ by making use of the relations
\begin{align} 
\extd\xi_c & = e^{Q_c}\left(w^{\prime}_c\extd D_c-2 \Lambda h^{\prime}_c\extd D_c-2\extd M-2 h_c\extd \Lambda\right)  \nonumber \\
& \quad + U_c\xi_c\extd D_c \label{eq:dxic} \\ \label{eq:dM}
\extd M & = T \extd S-h_h \extd\Lambda\,.
\end{align}
where the subscript $h$ indicates evaluation at $X_h$. Thus, one arrives at
\begin{equation}
\extd H_c(S,\,\Lambda,\,D_c) = T_c\extd S-\Theta_c \extd \Lambda-\psi_c\extd D_c \label{SSfirst}\,,
\end{equation}
with the temperature $T_c$ from \eqref{eq:lalapetz}, the dilaton chemical potential $\psi_c$ calculated in \eqref{SSpsi}, and the thermodynamical volume $\Theta_c$ determined in \eqref{Theta}. This is the quasi-local form of the first law of black hole thermodynamics in dilaton gravity with the cosmological constant treated as a thermodynamic variable (for a discussion of the difference between the quasi-local and asymptotic forms of the first law, see \cite{Grumiller:2007ju}). The three conjugate pairs of variables are 1.~local temperature $T_c$ and entropy $S$, 2.~thermodynamical volume $\Theta_c$ and (sign-reversed) cosmological constant $-\Lambda$, and 3.~dilaton charge $D_c$ and dilaton chemical potential $\psi_c$. \par
When written on the ``matter side'' of Einstein's equations, the cosmological constant acts like a perfect fluid with pressure $p=-\frac{\Lambda}{8 \pi G_{d+1}}$.
Thus, if we identify the cosmological constant as a negative pressure and denote the volume as $V_c=8\pi G_{d+1}\Theta_c$ (for intrinsically 2d models we have $p=-\Lambda, V_c=\Theta_c$ due to our conventions), then the first law above takes the familiar form
\eq{
\extd H_c(S,\,p,\,\dots) = T_c \extd S + V_c \extd p  + \textrm{other\;work\;terms}\,.
}{eq:familiar}
Similarly, the first law for the Gibbs free energy reads in this notation
\eq{
\extd G_c(T,\,P,\,\dots) = - S \extd T_c + V_c \extd p  + \textrm{other\;work\;terms}
}{eq:familiartoo}
while the one for internal energy is given by
\eq{
\extd E_c(S,\,V_c,\,\dots) = T_c \extd S - p \extd V_c  + \textrm{other\;work\;terms}
}{eq:toofamiliar}
and the one for Helmholtz free energy by
\eq{
\extd F_c(T_c,\,V_c,\,\dots) = - S \extd T_c - p \extd V_c  + \textrm{other\;work\;terms}
}{eq:stillfamiliar}
Thus, standard thermodynamics is recovered. The asymptotic form of the first law is derived in the next section, and an equivalent microcanonical analysis is presented in section \ref{se:7c}.

For vanishing cosmological constant (\ref{SSfirst}) reduces to the first law derived in \cite{Grumiller:2007ju}.  We stress again that, in the framework of 2d dilaton gravity, treating the cosmological constant as a thermodynamical variable requires no further assumptions in this setup, since it is just a charge with a specific coupling to the dilaton field. The first law for dilaton gravity coupled to a generic Maxwell field essentially looks the same. The only differences lie in the definition of the coupling function and the resulting function $h(X)$, which always lead to confinement for a cosmological constant, and the extensivity properties discussed in the beginning of this section.

\subsection{Asymptotic form of the first law}\label{subsec:AsymptoticFirstLaw}

The conserved quantity $M$, the mass of the black hole, is related to the asymptotic limit of the proper enthalpy $H_c$ via \begin{equation}
\lim_{X_c\rightarrow \infty}\sqrt{\xi_c}\,H_c= M\,. \label{Masslimit} 
\end{equation}
One can readily use the condition $\xi(X_h) = 0$ and the result \eqref{eq:S} for the entropy to derive the first law \eqref{eq:dM} for $M$. But this same result also follows from \eqref{Masslimit} and the quasi-local form of the first law \eqref{SSfirst}. Differentiating the quantity $\sqrt{\xi_c}\,H_c$ gives
\begin{align} \nonumber
	\extd\,(\sqrt{\xi_c}\,H_c) = & \,\, \sqrt{\xi_c}\,T_c \extd S - \sqrt{\xi_c}\,\Theta_c \extd \Lambda - \sqrt{\xi_c}\,\psi_c \extd D_c \\ 
	& + e^{-Q_c}\,\big(\sqrt{\xi_0} - \sqrt{\xi_c}\,\big)\,\frac{1}{2\,\sqrt{\xi_c}} \extd \xi_c ~.
\end{align}
Using \eqref{eq:dxic} for $\extd \xi_c$ and taking the $X_c \to \infty$ limit yields
\begin{equation}
	\lim_{X_c \to \infty} \extd\, (\sqrt{\xi_c}\,H_c) = \extd M = T \extd S +h_{h} \extd\,(-\Lambda)
\end{equation}
Thus, in the asymptotic form of the first law the quantity $\Theta_c \extd \Lambda$ is replaced by $h_{h} \extd \Lambda$, which identifies $h_{h}$ as the (asymptotic) thermodynamical volume $V$ of the black hole spacetime. Note that the correct expression for $V$ must be obtained from the first law; the $X_c \to \infty$ limit of $\sqrt{\xi_c}\,\Theta_c$ yields additional finite terms besides $h_{h}$.

\subsection{Generalization to charged black holes}\label{subsec:Generalization}

We generalize now equation (\ref{SSfirst}) for systems with additional $U(1)$ charges, like Reissner--Nordstr\"om or BTZ black holes. This introduces an additional term in the Killing norm for each charge
\begin{equation}\label{eq:ChargedKillingNorm}
\xi(X)=e^{Q(X)}\Big(w(X)-2M-2\Lambda h^\Lambda(X)+\sum_i\tfrac{q^2_i}{4}h^i(X)\Big)
\end{equation}
where the sum runs over the number of additional $U(1)$ fields. Here, the superscript $\Lambda$ indicates that the function $h^\Lambda(X)$ is associated with the cosmological constant coupling function (\ref{hcosmo}). The coupling functions $f^i(X)$ for the $U(1)$ gauge fields are left unspecified, subject only to the assumption that the $h^{i}(X)$ terms do not dominate the asymptotic behavior of \eqref{eq:ChargedKillingNorm}. 

As before, we want the ground state to be asymptotic AdS. Therefore we choose $\xi_0(X)$ to be the state with $M=q_{i}=0$. 
The first law for the Gibbs free energy is thus generalized to
\eq{
\extd G_c= -S \extd T_c +\sum_i\Phi_c^i\extd q_i - \Theta_c \extd \Lambda -\psi_c\extd D_c
}{chargefirst}
with $\Phi_c^i$ the proper electrostatic potential for the $i^{\text{th}}$ $U(1)$ field
\eq{
\Phi_c^i = \frac{\partial G_{c}}{\partial q_{i}} \bigg|_{T_c,\Lambda,D_c} = -\frac{q_i}{4}\,\frac{h_c^i-h_h^i}{\sqrt{\xi_c}}\,.
}{eq:phis}
Again, the subscripts $h$ and $c$ stand for evaluation at horizon and cavity wall, respectively. The addition of the new charges affects both the Killing norm \eqref{eq:ChargedKillingNorm} and the locus $X_h$ of the horizon, but it leaves the expressions \eqref{eq:S} for the entropy $S$ and \eqref{Theta} for the thermodynamical volume $\Theta_c$ unchanged. 
The dilaton chemical potential, however, acquires a new term
\begin{multline}
\psi_c=-\frac{\partial G_c}{\partial D_c}\Big|_{T_c,\Lambda,q_i} =-\frac{1}{2}U_ce^{-Q_c}\left(\sqrt{\xi_c}-\sqrt{\xi_0}\right)+\\
\left(\frac{1}{2}w^{\prime}_c-\Lambda h^{\prime}_c\right)\left(\frac{1}{\sqrt{\xi_c}}-\frac{1}{\sqrt{\xi_0}}\right) + \sum_{i}\frac{q_{i}^{2}}{8}\,\frac{h^{i}_{c}}{\sqrt{\xi_c}}\,. \label{SSpsi2}
\end{multline}


Generalizations to non-Abelian gauge fields are straightforward as well, but will not be discussed in detail in the present work. The main change as compared to the Abelian case is that the conserved charges $q_i$ are replaced by the conserved Casimirs of the corresponding gauge algebra. For each Casimir $C_i$ one can introduce a separate coupling function $f^i(X)$, exactly as above. For instance, in the case of $su(2)$ one would have only the quadratic Casimir as conserved quantity from the gauge sector, while for $su(3)$ one would have quadratic and cubic Casimirs, which could have separate coupling functions to the dilaton field. 

In the next section we address further developments of black hole thermodynamics in the presence of a state-dependent cosmological constant $\Lambda$. Those results also generalize to charged black holes in the same way as described above and to black holes with a confining $U(1)$ charge as described in section \ref{se:confinement}.

\section{Further developments}\label{se:7}

In this section we proceed with some further developments. In section \ref{se:7a} we present an alternative confinement condition that arises in some applications and violates our assumption \eqref{eq:con1}. In section \ref{se:7b} we unravel the presence of a Schottky anomaly in the specific heat and explain its physical origin. In section \ref{se:7c} we conclude with a microcanonical analysis thereby recovering in a simple way results from the canonical analysis in the limit when the cavity wall is removed to infinity.

\subsection{Alternative confinement condition}\label{se:7a}

The confinement condition \eqref{eq:con2} holds assuming that also the assumption \eqref{eq:con1} is true. If instead of \eqref{eq:con1} the function $w$ obeys
\eq{
\lim_{X\to\infty} w(X) = 0
}{eq:con1alt}
then the confinement condition \eqref{eq:con2} is replaced by 
\eq{
\lim_{X\to\infty} |h(X)| = \infty\,. 
}{eq:con2alt}
The whole discussion in section \ref{se:3.3} remains valid, in particular the condition \eqref{eq:Ures}.
We shall discuss two examples where the conditions \eqref{eq:con1alt}, \eqref{eq:con2alt} hold in section \ref{se:5}.

\subsection{Schottky anomaly}\label{se:7b}

The Schottky anomaly in the specific heat is the phenomenological observation that specific heat can have a maximum and decrease monotonically at sufficiently large temperatures. The attribute `anomaly' is justified because usually the specific heat increases with temperature (or remains constant) in standard condensed matter systems. Typically, the Schottky anomaly arises in systems with a limited number of energy levels, see e.g.~\cite{Kittel:1976in,Mohn:2006in}.

We argue now that generically we have a Schottky anomaly in the specific heat as long as we are working with a finite cut-off $X_c$. The specific heat (at constant dilaton charge $D_c$ and constant $\Lambda<0$) is given by
\eq{
C_c=T_c\,\frac{\partial S}{\partial T_c}\Big|_{D_c,\,\Lambda} = \frac{2\pi\,w_h^\prime-4\pi\,\Lambda h_h^\prime}{w_h''-2\Lambda h_h''+ \frac{(w_h^\prime-2\Lambda h_h^\prime)^2}{2(w_c-2M-2\Lambda h_c)}}
}{eq:C}
where the mass $M$ is determined as function of temperature $T_c$, cosmological constant $\Lambda$ and dilaton charge $D_c$ from the relation
\eq{
T_c = \frac{w_h^\prime-2\Lambda h_h^\prime}{4\pi\,\sqrt{e^{Q_c}\big(w_c-2M-2\Lambda h_c\big)}}\,.
}{eq:Msol}

For small black holes, $X_h\ll X_c$, the behavior of specific heat depends on the particular model. Either specific heat tends to zero from above as the black hole shrinks, or there is a Hawking--Page phase transition, in which case specific heat could have a pole at some critical temperature and become negative below that temperature. So there is no universal behavior of specific heat for small black holes.

For large black holes, $X_h = X_c (1-\epsilon)$ with $\epsilon \ll 1$ we expand in powers of $\epsilon$ and obtain
\eq{
C_c=\epsilon\, 4\pi X_h+{\cal O}(\epsilon^2)
}{eq:CSchottky}
%
This quantity is non-negative for positive $\epsilon$.
Thus, specific heat tends to zero from above as the horizon approaches the cut-off surface. At the same time, temperature is monotonically increasing in this limit,
\eq{
T_c=\frac{1}{\sqrt{\epsilon}} \, \frac{\sqrt{w^{\prime}_h-2\Lambda h_h^{\prime}}}{4\pi \sqrt{e^{Q_c} X_h}}+{\cal O}(\sqrt{\epsilon})\, ,
}{eq:TSchottky}
since the Tolman factor diverges if the black hole horizon coincides with the cut-off surface due to an infinite blueshift. Parametrically we have the relation
\eq{
C_c = \frac{N^2+{\cal O}(\epsilon)}{T_c^2}
}{eq:CTSchottky}
where $N^2$ is some numerical factor given by 
\begin{equation}
N^2=\frac{w^{\prime}_c-2\Lambda h^{\prime}_c}{4\pi e^{Q_c}}
\end{equation}
 that is independent from $\epsilon$. The $1/T_c^2$ behavior of specific heat in the high-temperature limit is typical for a non-interacting spin system (``ideal paramagnet'') in an external field, see e.g. \cite{Mohn:2006in}.

In conclusion, specific heat decreases strictly monotonically as a function of temperature for black holes whose size approaches the cut-off surface. If specific heat is monotonically increasing at some lower temperature, by continuity specific heat must have an extremum. This is the essence of the Schottky anomaly for the specific heat.

In fact, it is simple to understand the physical origin of our Schottky anomaly. Namely, the large growth of temperature is not associated with a large growth of states or a large growth of the black hole mass, because there is literally no room left for additional states as the black horizon is already close to the cut-off surface. Instead, the growth of temperature is almost entirely due to increased blueshifts. This is why the specific heat is decreasing, as it is the case for the Schottky anomaly.

The Schottky anomaly above exists for arbitrary models compatible with our assumptions, as long as we work at finite cut-off $X_c$. If one would like to have a Schottky anomaly that persists as the cut-off is removed to infinity then one would need to introduce functions $w$ and $h$ that are chosen such that the specific heat for $X_c\to\infty$,
\eq{
C_\infty = \lim_{X_c\to\infty} C_c = 2\pi\,\frac{w_h^\prime-2\Lambda h_h^\prime}{w_h''-2\Lambda h_h''}
}{eq:Cinfty}
is positive and has an extremum if expressed as a function of $T$. None of the models that we are going to discuss as examples has this property. It could be interesting to construct such examples in order to model Schottky anomalies in an AdS/CFT context.

\subsection{Microcanonical analysis}\label{se:7c}

While so far we have worked in the canonical ensemble (or related ensembles), our main result \eqref{eq:intro1}-\eqref{eq:intro2} can also be applied to microcanonical thermodynamics. In that case the cut-off, which is needed to ensure the existence of the canonical ensemble in most models, is not introduced. As a result there is no dilaton charge or dilaton chemical potential, and Tolman factors do not appear.

Let us start by directly deriving a microcanonical first law.
By analogy to \cite{Grumiller:2007ju} we formulate first `Smarr's law', i.e., a relation between mass parameter $M$ and other parameters (cosmological constant $\Lambda$ and value of the dilaton at the horizon $X_h$) 
\eq{
M = \tfrac12\,w(X_h)-\Lambda h(X_h) \,.
}{eq:mc3}
Differentiating this relation obtains
\eq{
\extd M = \big(\tfrac12\,w_h^\prime-\Lambda h_h^\prime\big)\,\extd X_h - h_h\,\extd \Lambda\,,
}{eq:mc2}
where the subscript $h$ denotes evaluation at the horizon.
With the Hawking-temperature $T=\beta^{-1}$ from \eqref{eq:beta}, the Bekenstein--Hawking entropy $S$ from \eqref{eq:S}, the pressure $p=-\Lambda/(8\pi G_{d+1})$, and the volume $V=8\pi G_{d+1}h_h$ [see \eqref{hcosmo}] this yields the microcanonical first law [see \eqref{eq:dM}]
\eq{
\extd M(S,\,p) = T \extd S + V \extd p\,.
}{eq:mc5}
This is the same result obtained via an asymptotic analysis in section \ref{subsec:AsymptoticFirstLaw}.


In the next section we consider various examples of canonical and micro-canonical black hole thermodynamics in the presence of a state-dependent cosmological constant $\Lambda$.

\section{Examples}\label{se:5}

In this section we present three examples. The purpose of the first two examples is to confirm that our general results in sections \ref{se:2}, \ref{se:3}, \ref{se:4} and \ref{se:7} are correct, so we focus on recovering known results in the language of 2d dilaton gravity.  We stress here the generality of our results, which can be applied to any 2d dilaton gravity theory, subject to the conditions \eqref{eq:con1} and \eqref{eq:con2} [or the alternative conditions \eqref{eq:con1alt} and \eqref{eq:con2alt}] with \eqref{eq:important} on the model dependent functions $U(X)$ and $V(X)$. As the last example we discuss one of the simplest intrinsically 2d models whose solutions asymptote to AdS and whose potentials obey the alternative confinement conditions \eqref{eq:con1alt} and \eqref{eq:con2alt} [again with \eqref{eq:important}].

\subsection{AdS-Schwarzschild--Tangherlini}

Since AdS-Schwarzschild--Tangherlini is an effectively 2d spacetime [fibered by $(d\!-\!1)$-spheres] it can be described in the framework of 2d dilaton gravity. For a thorough treatment of the thermodynamics of this spacetime see e.g.~\cite{Brown:1994fk}. We will make use of the prescription for spherical reduction of higher dimensional black holes given in \cite{Grumiller:2007ju} and take into account the proportionality factor omitted in \eqref{eq:UVsrg}, which involves Newton's constant in $d\!+\!1$ dimensions, $G_{d+1}$.
\begin{align}
U(X)&=-\frac{d-2}{(d-1)X}\\
V(X)&=-\frac{1}{2}(d-1)(d-2)\Upsilon^{\frac{2}{d-1}}X^{1-\frac{2}{d-1}}.
\end{align}
Here $\Upsilon=\frac{A_{d-1}}{8\pi G_{d+1}}$, where $A_{d-1}$ denotes again the solid angle subtended by the $(d\!-\!1)$-dimensional unit sphere. We assume $1<d<\infty$.

According to our general discussion we add now a $U(1)$ gauge field with coupling function \eqref{eq:important}.
The functions $w(X)$, $e^{Q(X)}$ and $h(X)$ are easily calculated to yield
\begin{align}
w(X)&=(d-1)\Upsilon^{\frac{1}{d-1}}X^{\frac{d-2}{d-1}}\\
e^{Q(X)}&=\frac{1}{d-1}\Upsilon^{\frac{1}{1-d}}X^{\frac{2-d}{d-1}}\\
h(X)&=\frac{1}{d}\Upsilon^{\frac{1}{1-d}}X^{\frac{d}{d-1}}
\end{align}

We recover the first law  \eqref{SSfirst} for AdS-Schwarzschild--Tangherlini black holes with variable cosmological constant, with the volume
\eq{   
\Theta_c=\frac{1}{d}\Upsilon^{\frac{1}{1-d}}\Big(\frac{X_c^{\frac{d}{d-1}}}{\sqrt{\xi_0}}-\frac{X_c^{\frac{d}{d-1}}}{\sqrt{\xi_c}}+\frac{X_h^{\frac{d}{d-1}}}{\sqrt{\xi_c}}\Big)\,.
}{eq:ASTv}
The dilaton field is related to the radial coordinate $r$ of $(d\!+\!1)$-dimensional AdS-Schwarzschild via
\begin{equation}
X(r)=\Upsilon r^{d-1}.
\end{equation}
From the results above the known relations for temperature and surface pressure follow (see \cite{Brown:1994fk}). 


If written on the right side of Einstein's equation the cosmological constant can be regarded as a negative pressure
$p=-\Lambda/(8\pi G_{d+1})$, concurrent with our general discussion in section \ref{se:4b}.
If the cavity is removed, $X_c\to\infty$, then the asymptotic form of the first law given in section \ref{subsec:AsymptoticFirstLaw} is
\begin{equation}
\extd M(S,\,p)=T\extd S + V\extd p \label{enthalpy}
\end{equation}
where $V = 8\pi G_{d+1} h_{h}$ is the volume of a $(d\!-\!1)$-sphere with radius $r_{h}$
\begin{equation}
	V = \frac{1}{d}\,A_{d-1}\,r_{h}^{\,d} ~.
\end{equation}
Thus, the mass should be regarded as enthalpy in AdS-space. This result was discussed already in \cite{Dolan:2010ha}, \cite{Dolan:2012jh,Dolan:2011xt,Cvetic:2010jb,Kastor:2009wy}. 

\subsection{BTZ black holes}

BTZ black holes are solutions of Einstein gravity with negative cosmological constant in three dimensions \cite{Banados:1992wn}, \cite{Banados:1992gq}. In \cite{Achucarro:1993fd}, a Kaluza--Klein reduction to 2d of BTZ black holes was presented. However, in this treatment the angular momentum $J$ enters as a parameter in the action rather than emerging as a constant of motion. In the same way that the cosmological constant can be promoted to a constant of motion by integrating in a $U(1)$ field, a nonzero angular momentum of BTZ can be treated by introducing an additional $U(1)$ field with a specific coupling and charge $q=J$ \cite{Grumiller:2007ju}.

The following functions are used to model BTZ in 2d dilaton gravity:
\begin{align} \nonumber
e^{Q(X)}&=4 G_3 &  w(X)&=0 \\ \label{BTZfunc} 
f^{\Lambda}(X)&=\frac{1}{X} &  f^{J}(X)&= -2 G_{3}^{\,2} X^{3}\\ \nonumber
 h^\Lambda(X)&=2 G_{3} X^2 &  h^J(X)& =\frac{1}{G_3 X^2} 
\end{align}
The BTZ black hole is thus a model satisfying the generalized confinement condition given in section \ref{se:7a}.

With the functions (\ref{BTZfunc}) in the action, the analysis of section \ref{subsec:Generalization} yields the first law for BTZ black holes with the cosmological constant as thermodynamic variable. In terms of the (proper) enthalpy the first law reads
\eq{
\extd H_c=T_c\extd S+\Omega_c\extd J-\Theta_c\extd\Lambda-\psi_c\extd D_c\,,
}{eq:1stBTZ}
with 
\begin{align}\label{BTZTc}
T_c&=\frac{2 G_3}{\pi \ell^{2} \sqrt{\xi_c}}\bigg(X_h - \frac{1}{X_h^{3}}\,\bigg(\frac{\ell J}{4 G_3}\bigg)^2\,\bigg) \\ \label{BTZOmegac}
\Omega_c&=\frac{J}{4 G_3 \sqrt{\xi_c}}\left(\frac{1}{X_h^2}-\frac{1}{X_c^2}\right) \\ \label{BTZpsic}
\psi_c&=\frac{4 G_3}{\sqrt{\xi_c}}\,\left(\frac{1}{\ell}\,H_{c} - \frac{1}{X_{c}^{3}}\bigg(\frac{J}{4 G_3}\bigg)^{2}\,\right) \\ \label{BTZThetac}
\Theta_c&= \left(\frac{h^{\Lambda}_{\,c}}{\sqrt{\xi_0}}-\frac{h^{\Lambda}_{\,c}}{\sqrt{\xi_c}}+\frac{h^{\Lambda}_{\,h}}{\sqrt{\xi_c}}\right)\,,
\end{align}
where we have introduced the AdS radius $\ell$ related to $\Lambda$ in three dimensions via $\Lambda=-\frac{1}{\ell^2}$. The dilaton is related to the standard AdS radial coordinate $r$ by $X = r/(4 G_3)$. Using this to rewrite the thermodynamic quantities in terms of the inner horizon $r_{-}$ and outer horizon $r_{+}$, given by
\begin{gather}
	r_{\pm} = \ell\,\sqrt{4 G_3 M}\,\sqrt{1 \pm \sqrt{1 - \frac{J^{2}}{M^{2}\ell^{2}}}} ~,
\end{gather}
gives the thermodynamic quantities \eqref{BTZTc}-\eqref{BTZThetac} in their more familiar forms.

Again, we express the cosmological constant in terms of the associated pressure $p=-\frac{\Lambda}{8\pi G_3}$. Accounting for Tolman factors, using the relation (\ref{Masslimit}), and taking the limit $X_c\to\infty$ one recovers the analogue of \eqref{enthalpy} for BTZ black holes, which was already presented in \cite{Dolan:2012jh}:
\eq{
\extd M(S,\,J,\,p) = T\extd S + \Omega\extd J + \pi r_+^2\extd p
}{eq:2ndBTZ}
The last term again contains the geometric volume $V=\pi r_+^2$ of the black hole, see \cite{Grumiller:2005zk}.

\subsection{Jackiw--Teitelboim model}

The Jackiw--Teitelboim model \cite{Teitelboim:1984,Jackiw:1984} has vanishing kinetic potential, $U(X)=0$, and linear dilaton self-interaction potential, $V(X)=\Lambda X$. Like in our general discussion we convert the parameter $\Lambda$ of the action into a constant of motion by integrating in a Maxwell field.

In our conventions the relevant functions read
\eq{
Q(X)=w(X)=0\qquad h(X)=\frac{X^2}{2}\,.
}{eq:JT}
The Killing norm \eqref{chargedKill} simplifies to
\eq{
\xi =(-\Lambda)X^2 - 2M
}{eq:JTkill}
so that the Killing horizon for negative $\Lambda=-1/\ell^2$ is located at
\eq{
X_h = \ell\,\sqrt{2M}\,.
}{eq:JTXh}

The confinement conditions \eqref{eq:con1alt} and \eqref{eq:con2alt} hold, so the gauge field is confining.
\eq{
A_\tau(X) = -\frac{\sqrt{2}}{4\ell}\,\big(X^2-X_h^2\big) + A_\tau(X_h)\qquad A_r=0
}{eq:JTA}
Thus, the Jackiw--Teitelboim model is the simplest example of the class of models that satisfy the generalized confinement condition presented in section \ref{se:7a}.

The Hawking temperature \eqref{eq:beta} yields
\eq{
T= \frac{\xi^\prime}{4\pi}\Big|_{X=X_h} \!\!\!\!=  \frac{\sqrt{2M}}{2\pi\ell}
}{eq:JTT}
so that the Tolman temperature \eqref{eq:lalapetz} is given by
\eq{
T_c = \frac{\sqrt{2M}}{2\pi\sqrt{X_c^2-2\ell^2M}}
}{eq:JTTc}
which leads to a unique value for the mass as a function of $T_c$, $\Lambda$, and $D_c=X_c$:
\eq{
M(T_c,\,\Lambda,\,D_c)=\frac{2\pi^2T_c^2 \Lambda D_c^2}{\Lambda-4\pi^2T_c^2}
}{eq:JTM}

The Gibbs free energy reads
\begin{multline}
G_c(T_c,\,\Lambda,\,D_c) = D_c\,\big(\sqrt{-\La} -\sqrt{4\pi^2T_c^2-\Lambda} \,\big)\,.
\label{eq:JTG}
\end{multline}
From the Gibbs free energy we derive
\begin{align}
S &= -\frac{\partial G_c}{\partial T_c}\Big|_{\Lambda,\,D_c} \!\!\!\!= \frac{4\pi^2 T_c D_c}{\sqrt{4\pi^2T_c^2-\Lambda}} = 2\pi X_h \\
\Theta_c &= -\frac{\partial G_c}{\partial\Lambda}\Big|_{T_c,\,D_c} \!\!\!\! = \frac{D_c}{2}\,\Big(\frac{1}{\sqrt{-\La}}-\frac{1}{\sqrt{4\pi^2T_c^2-\La}}\Big)\\
\psi_c &= -\frac{\partial G_c}{\partial D_c}\Big|_{T_c,\,\Lambda} \!\!\!\! = \sqrt{4\pi^2T_c^2-\Lambda} -\sqrt{-\La}\,.
\end{align}
As it must be, the first law holds.
\eq{
\extd G_c = -S\extd T_c-\Theta_c\extd\Lambda -\psi_c\extd D_c
}{eq:JT1st}

For small temperatures, $T_c^4\ll \Lambda^2$, entropy vanishes linearly, while volume and dilaton chemical potential vanish quadratically in $T_c$. 
For large temperatures, $T_c^4\gg \Lambda^2$, entropy and volume approach constant values, while the dilaton chemical potential diverges linearly in $T_c$.

Other thermodynamical quantities of interest can be calculated straightforwardly. As an example we calculate the specific heat at constant $\Lambda$ and constant dilaton charge and find
\eq{
C_c\big|_{\Lambda,\,D_c} \!\! = T_c \,\frac{\partial S}{\partial T_c}\Big|_{\Lambda,\,D_c} \!\!\!\!= \frac{4\pi^2 T_c(-\La) D_c}{(4\pi^2T_c^2-\La)^{3/2}} \geq 0\,.
}{eq:JTC}
Thus, specific heat at constant $\Lambda$ and constant dilaton charge is non-negative for the Jackiw--Teitelboim model and vanishes linearly at low temperatures, just like a free Fermi gas. The high-temperature behavior, which is valid in the limit of large black holes, $X_h^2 \gg X_c^2-X_h^2$, is given by 
\begin{equation}
C_c \sim \frac{D_c (-\Lambda)}{2\pi T_c^2}+{\cal O}(T_c^{-4})
\end{equation}
in accordance with equation \eqref{eq:CTSchottky} and thus exhibits the Schottky anomaly predicted from our general discussion in section \ref{se:7b}.

\acknowledgments

DG and JS were supported by the START project Y~435-N16 of the Austrian Science Fund (FWF) and the FWF projects I~952-N16 and I~1030-N27. 
RM thanks KITP, and partial support from NSF grant PHY11-25915, for hospitality during the completion of this work.


\providecommand{\href}[2]{#2}\begingroup\raggedright\endgroup

\end{document}